\documentclass[12pt]{article}

\usepackage{natbib}
 \bibpunct[, ]{(}{)}{,}{a}{}{,}%
\usepackage[utf8]{inputenc}
\usepackage{graphicx}
\usepackage[ruled,vlined,linesnumbered]{algorithm2e}
\usepackage{amssymb}
\usepackage{comment}
\usepackage{amsmath}
\usepackage{textcomp}
\usepackage{lscape}
\usepackage[ruled,vlined]{algorithm2e}
\usepackage{endnotes}
\let\footnote=\endnote

\usepackage{subcaption}
\usepackage{amsmath}
\graphicspath{ {./images/} }
\usepackage{setspace}
\usepackage{algcompatible}
\usepackage{amssymb,amsmath,amsfonts,eurosym,geometry,ulem,graphicx,caption,color,setspace,sectsty,comment,footmisc,caption,pdflscape,array,subcaption}
\usepackage{authblk}

\usepackage{titlesec}

\title{Solving Multi-Period Financial Planning Models: Combining Monte Carlo Tree Search and Neural Networks}

\begin{document}

\author[1]{Afşar Onat Aydınhan}
\author[1]{Xiaoyue Li}
\author[1]{John M. Mulvey}

\affil[1]{Department of Operations Research and Financial Engineering, Princeton University}

\maketitle

\begin{abstract}
 This paper introduces the MCTS algorithm to the financial world and focuses on solving significant multi-period financial planning models by combining a Monte Carlo Tree Search algorithm with a deep neural network. The MCTS provides an advanced start for the neural network so that the combined method outperforms either approach alone, yielding competitive results.  Several innovations improve the computations, including a variant of the upper confidence bound applied to trees (UTC) and a special lookup search. We compare the two-step algorithm with employing dynamic programs/neural networks. Both approaches solve regime switching models with 50-time steps and transaction costs with twelve assets. Heretofore, these problems have been outside the range of solvable optimization models via traditional algorithms.   
\end{abstract}


\section{Introduction}

Monte Carlo tree search (MCTS) is a heuristic based multi-purpose search algorithm which searches for optimal decision by combining the concepts of tree search and reinforcement learning. The MCTS algorithm dates back to 2006, when a Computer Go program won the 10th KGS computer-Go tournament (\citeauthor{coulom_2007} \citeyear{coulom_2007}). Since then, MCTS has been applied in a variety of fields, from security evaluation methodology of image-based biometrics authentication systems (\citeauthor{tanabe_yoshizoe_imai_2009} \citeyear{tanabe_yoshizoe_imai_2009}) to murder mystery generation (\citeauthor{jaschek_beckmann_garcia_raffe_2019} \citeyear{jaschek_beckmann_garcia_raffe_2019}). The algorithm has gained popularity over the past several years, especially due to the success of DeepMind’s AlphaZero, a program that beats the reigning world champion programs in games like chess, shogi and go (\citeauthor{Silver2017MasteringCA} \citeyear{Silver2017MasteringCA}) .

The original MCTS algorithm is mainly designed to simulate and move forward in time in a multi-period decision process, that is the algorithm slowly builds a search tree depending on the simulation results where the depth of a node in the tree represents how far away into the future it is. However, depending on the problem, the convergence can be prohibitive. In our LMCTS (Lookup Monte Carlo Tree Search) algorithm, we aim to remedy this by combining the strengths of the MCTS algorithm and dynamic programming. We start solving our multi-period problem from the end and store the results of the MCTS algorithm on a lookup table. Then we move backward in time and solve a bigger problem by using the results of the lookup table as our rollout policy in the simulation phase of the MCTS algorithm. This niche algorithm can only be employed in a problem where the number of state-action pairs are tractable, otherwise the time it would take to calculate the lookup table and its size would explode. The advantage of the algorithm over its counterpart, dynamic programming, is its flexibility thanks to its simulative nature. The new algorithm is able to handle complex problem definitions, objective functions and constraints on the state space.

MCTS was originally designed as to deal with problems that have finite discrete action spaces.  Different policies are available within the algorithm, but most famous is the classical UCT method (\citeauthor{kocsis_szepesvari_2006} \citeyear{kocsis_szepesvari_2006}) which tackles the exploration-exploitation dilemma that occurs in problems with stochastic outcomes. Multiple variants have been developed to extend the MCTS algorithm to continuous action spaces. Progressive widening (\citeauthor{coulom_2007} \citeyear{coulom_2007}) or progressive pruning (\citeauthor{chaslot} \citeyear{chaslot}) adopts a growing set of discrete actions to handle continuous action spaces. cRAVE (\citeauthor{couetoux_hoock_sokolovska_teytaud_bonnard_2011} \citeyear{couetoux_hoock_sokolovska_teytaud_bonnard_2011})   uses a similar idea applied to the RAVE algorithm. HOOT (\citeauthor{Mansley2011SampleBasedPF} \citeyear{Mansley2011SampleBasedPF}) applies the HOO (\citeauthor{bubeck} \citeyear{bubeck}) algorithm in UCT instead of UCB. However, for our problem setting, we have chosen KR-UCT (\citeauthor{Yee2016MonteCT} \citeyear{Yee2016MonteCT}) to be the most suitable variant of the MCTS that handles continuous action spaces. It is advised for the reader to check the aforementioned paper which analyzes the KR-UCT algorithm and compares it empirically to other algorithms for continuous action spaces. Although the algorithm is mainly developed for “actions with execution uncertainty” it works similarly for our financial decision model under uncertainty (In contrast, a problem setting without any action execution or outcome uncertainty is chess). The distinguishing traits of KR-UCT are explained in (\citeauthor{Yee2016MonteCT} \citeyear{Yee2016MonteCT}) as follows: a) information sharing between all actions under consideration, b) identification of actions outside of the initial candidates for further exploration, c) selection of actions outside of the candidate set. The benefits of these traits become clearer once the problem setting is established in the next section. In short, due to time constraints, we may not have adequate computational resources to run enough simulations at each time step. Thus, the algorithm benefits from information sharing between all actions under consideration. And since the action space under consideration is massive, identification of actions outside of the initial candidates and selection of actions outside of the candidate set are traits which are extremely valuable to us.

A multi-period asset allocation problem is the focus of this study. Multi-period problems enjoys its farsightedness over single-period investment planning, where economic regime dynamics as well as investors' life cycle goals are better cooperated in the model. Elegant as single-period investment models, they fail to address the issues such as intermediate cash flows, transaction costs and trade-off between short-term versus long-term benefits. In the early days of financial mathematics, multi-period asset allocation problems are often solved with analytical solutions or by numerical methods like dynamic programs. When numerical methods are employed, the sizes of problems are usually constrained by computing powers. Though powerful in solving multi-stage optimization problems, vanilla dynamic program suffers the curse of dimensionality, a phenomenon that the required running time grows exponentially in the complexity of the problem. Modern algorithms are developed over the years to overcome the curse of dimensionality, Monte Carlo Tree Search and neural networks among the most known ones (\cite{alphago}).

The Merton portfolio problem is among the earliest and most famous inter-temporal investment problems. (\citeauthor{merton1969} \citeyear{merton1969}) proposes and finds a closed form solution for a continuous-time asset allocation problem where an investor maximizes her utility function for both finite and infinite lifetime. Since then, portfolio management over long horizon has been widely studied, and various extensions are made to the original Merton's problem to better describe the real economic environment. Literature finds that historical financial data have heavier left tails than normal distribution, and it is therefore not representative to model the returns with a symmetric distribution such as normal distribution. In particular, the returns form volatility clusters when crash happens, in which period asset returns behave substantially different to normal periods. (\citeauthor{Nystrup2018} \citeyear{Nystrup2018}) argue that investors benefit from dynamic strategies that weigh assets differently in crash versus in normal regime. In this paper, we follow (\citeauthor{li2021} \citeyear{li2021}) and analyze the trading strategies in a market switching between two possible regimes - normal and crash - with a Markov process. They present results for a 50-period 11-asset problem involving multiple regimes solved by a combined method with dynamic program and neural network, the size of which is untraceable with traditional methods alone. Here, we compare the performance and efficiency of several numerical methods, under various constraints on budgeting, transaction costs and utility functions. We will be tackling a 50-period stochastic optimization problem, which is quite an ambitious task to say the least.

The main contribution of the paper is to introduce the benefits of the MCTS algorithm to the financial portfolio optimization literature. To our knowledge, MCTS was not used to tackle a  multi-period financial portfolio optimization problem under a regime-switching framework before. In addition, we extend the vanilla  MCTS algorithm via novel version of it called LMCTS. We also employ a variant of the UCT algorithm called KR-UCT which is more suitable for the problem at hand due to reasons explained in the previous paragraphs. We solve our multi-period problem without transaction costs and use the solution of the LMCTS algorithm as a starting point for a neural network which handles the transaction cost of the problem. We compare the results with another neural network which uses a dynamic programming solution as a starting point.

The paper is organized as follows. Section 2 describes the underlying model. Here we define our general multi-period portfolio allocation problem which utilizes a Markov regime switching framework, and specify the utility functions and transaction costs that are used. Section 3 explains the methodologies used to tackle the problem at hand, which are MCTS, neural networks, and dynamic programming. Firstly, the Monte Carlo tree search algorithm and its four main steps (selection, expansion, simulation and backpropagation) are described. Next, the LMCTS algorithm, the KR-UCT function and our specific implementation of the Monte Carlo tree search for the problem is explained thoroughly. Then, general neural networks, recurrent neural networks and the way our neural networks are trained for the problem at hand are described. The section concludes with a brief discussion of dynamic programming. Section 4 presents our empirical results. The main goal is to compare the performances of starting algorithms for the recursive neural network, a Monte Carlo tree search solution and a dynamic programming solution as starting points respectively. We explore cases where we allow shorting and do not allow shorting separately. Also we observe algorithmic performance under different utility functions, namely probability of reaching a goal and terminal utility maximization. Section 5 contains the concluding remarks.

\section{Model}

In our multi-period portfolio optimization problem, we assume $n\geq 1$ risky assets and one risk-free asset in the market (the risk-free asset can be considered as cash) where the prices of risky assets
follow a set of correlated geometric Brownian motions processes. The parameters of the Brownian motions (the means and the covariance matrix) depend on the regime. The values of these parameters are inferred from historical values of real assets. A hidden Markov model is employed for the regimes, meaning that the investors do not know the regime that they are in, but they can infer the probability of being in a regime based on the asset returns they observe. The goal of the investor is to maximize their terminal utility and to do that she may rebalance her portfolio at the end of each period.

\subsection{The general multi-period portfolio allocation problem}
The general model is as follows:
\begin{align}
\begin{split}
\nonumber
    \underset{x_0,x_1,...,x_{T-1}\in\mathbb{R}^n}{\text{Maximize }}& \hspace{1cm} Utility[Z_1,Z_2,...]  \hspace{1cm}(1)\\
    \text{subject to} &\hspace{1cm} \bold{1}^T x_t = 1 \hspace{0.5cm} \forall t=0,...,T-1 \hspace{1cm}(2)\\
    &\hspace{1cm} W_{t}^{\rightarrow} = W_{t} (x_t^T(1+r_t)) \hspace{0.5cm} \forall t=0,...,T-1 \hspace{1cm}(3)\\
    &\hspace{1cm} x_{t}^\rightarrow = \frac{x_t\odot(1+r_t)}{x_t^T(1+r_t)} \hspace{0.5cm} \forall t=0,...,T-1 \hspace{1cm}(4)\\
    &\hspace{1cm} W_{t+1} = W_{t}^{\rightarrow} - C(W_{t}^{\rightarrow}; x_{t}^\rightarrow, x_{t+1})\hspace{0.5cm} \forall t=0,...,T-1 \hspace{1cm} (5)
\end{split}
\end{align}

\doublespacing

where \textit{T} is the number of periods, $x_0,x_1,...,x_{T-1}\in\mathbb{R}^n$ are the decision variable for the asset allocations at the beginning of each period, $\bold{1}\in\mathbb{R}^n$ is the vector of all ones, $W_t$ is the wealth at the beginning of period $t$, $W_t^\rightarrow$ is the wealth at the end of period $t$, $r_t\in\mathbb{R}^n$ is the vector of returns in period $t$, $\pi_t^\rightarrow\in\mathbb{R}^n$ is the allocation vector at the end of period $t$, $\odot$ is the element-wise multiplication operator, and $C(W; \pi^\rightarrow,\pi)$ is the dollar value of transaction and market impact costs when the allocation is rebalanced to $\pi^\rightarrow$ from $\pi$ with current wealth being $W$, and $F_t$ is the cash flow at time \textit{t}. We assume that the initial wealth, $W_0$, is given. The distribution of asset returns $r_t$ and the function of transaction cost $C(\cdot;\cdot,\cdot)$ depend on the problem at hand.

The objective (1) is to maximize a utility function which is most likely to be a function of the terminal wealth $W_T$, such as expected terminal wealth or probability of reaching a goal with the terminal wealth. Constraint (2) ensures that the total percentage allocation to assets is equal to 1. Equation (3) updates the wealth vector according to realized returns and asset allocations. Equation (4) updates the asset allocation vector according to the returns. Equation (5) calculates the wealth in the next period using wealth at the end of this period and the transaction cost.

\subsection{Specifications in our model}

\subsubsection{Utility Function}\hfill

The investor in our model will be aiming to maximize the expected value of her utility function, which is a function of the terminal wealth:

$$\underset{x_0 , x_1,..., x_{T-1}\in \mathbb{R}^n}{\textnormal{Maximize}} \mathbf{E}[U(W_T)] \hspace{1cm} (6)$$

The two different utility functions that we will be looking at are the CRRA(constant relative risk aversion) utility function:

$$ U_1(W) = \begin{cases}
  \frac{W^\gamma}{\gamma}, \hspace{1cm}  \gamma \neq 0      \hspace{1cm} (7) \\
  log(W), \hspace{1cm}  \gamma = 0    
\end{cases}$$

and the probability of reaching a goal, G:
$$ U_2(W) = \begin{cases}
  1, \hspace{1cm}  W \geq G              \hspace{1cm} (8) \\
  0, \hspace{1cm}  W < G 
\end{cases}$$ 

The CRRA utility function (7) and the non-convex probability of reaching a goal utility function (8) will be used to compare DP-DNN and MCTS-DNN methods.

\subsubsection{Regimes and Returns}\hfill

As mentioned, we assume a Markov regime switching framework. The $r_t$ in our model follows a set of correlated Geometric Brownian motions where the parameters are:

$$ r_t \sim \mathbb{N}(\mu_{S_t}, \Sigma_{S_t}), \hspace{1cm} S_t\in\{1,2,..,N\}  \hspace{1cm} (9)$$

$S_t$ represents the regime at time $t$. The returns on the risk-free rate $r_f$ are fixed, but the value may depend on the regime.

\subsubsection{Transaction Cost}\hfill
The transaction cost function $C(\cdot;\cdot,\cdot)$ can take any form. We assume a linear function in this paper.

\section{Methodology}

\subsection{Monte Carlo Tree Search}

\subsubsection{The original MCTS algorithm}\hfill 

Monte Carlo tree search (MCTS) is a general search algorithm for finding optimal solutions for problems that can be modeled as decision processes. This algorithm constitutes the core of AlphaZero, which actually defeated world champion Ke Jie in Go and top chess and shogi engines in chess and shogi respectively as well. This proves that Monte Carlo tree search algorithm indeed is an algorithm with an extraordinary potential(\citeauthor{fu} \citeyear{fu}). Four steps of the MCTS algorithm are: Selection, Expansion, Simulation and Backpropagation.

\begin{figure}
    \centering
    \includegraphics[width=1\textwidth]{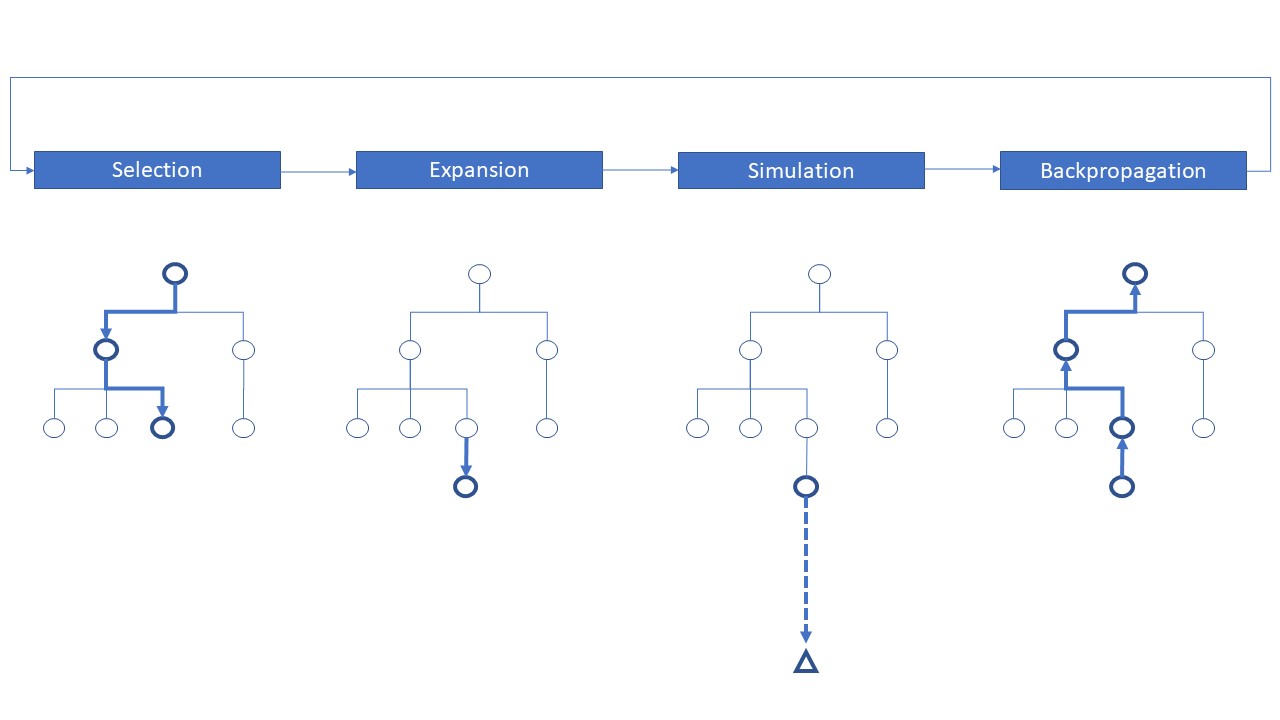}
    \caption{The primary steps of MCTS.}
    \label{fig:obj_goal5}
\end{figure}

\textbf{Selection:} The selection step corresponds to moving down the tree and selecting a node based on a tree policy. The tree policy is basically an evaluation function according to which the algorithm decides on which path to follow. Different tree policies can be adopted, but the most famous and common one is the Upper Confidence Bounds Applied to Trees (UCT)(\citeauthor{kocsis_szepesvari_2006} \citeyear{kocsis_szepesvari_2006}). UCT is a function of the mean return of the node and the number of times the node is visited, and with the help of a tunable constant, it allows the user to tackle the exploration-exploitation dilemma. The UCT function is as follows:

 $${v} + C \sqrt{\frac{log N }{n}}   \hspace{1cm} (10)$$
 
where ${v}$ is the mean reward of the node, $N$ is the total number of simulations done from the parent node, $n$ is the number of simulations done through the node at hand and $C$ is the tunable exploration parameter. This function is calculated for every child node of the parent node and the one with the highest function value is selected.

\textbf{Expansion:} Expansion is the creation of a new leaf node in the tree. The expansion can be made at a node which already has child nodes, but is not fully expanded (This means that not all possible actions have been taken at that node yet). Expansion can be made from a leaf node as well, in which case the newly created node would be the first child of the parent node. 

\textbf{Simulation:} Simulation is the process of simulating from the chosen/newly created node. This process is also called rolling out. Many different rollout policies can be used, such as the uniform random rollout policy in which the actions taken during the simulation phase are chosen at random with equal probability. 

\textbf{Backpropagation:} Backpropagation is the act of updating values of the nodes that were along the simulation path. After the simulation is over and a reward is obtained, backpropagation algorithm updates the value of all the nodes along the traversed path. The values that are of interest are the number of times the nodes are visited and the mean rewards of the simulations that the respective nodes participated in.

\subsubsection{LMCTS}\hfill

In the original MCTS algorithm the simulation phase is guided by what is called a rollout policy, which determines the actions taken during the simulations. We need to use a rollout policy that we do not yet know what the optimal action is in the future time steps. The hope is that the algorithm will still converge with enough simulations, regardless of how inaccurate the rollout policy is. There are different types of rollout policies. If there is no domain knowledge, the rollout policy generally implemented is the uniform random rollout policy. That is, we choose our action randomly with equal chances for each possible action at each node during the rollout phase. However, if we possess domain knowledge that can guide us during the rollout phase, we aim to outperform the uniform random rollout policy (\citeauthor{James2017AnAO} \citeyear{James2017AnAO}).

If a superior rollout policy improves performance, the best performance should be obtained using the ultimate best rollout policy, which is the optimal policy at each time step. This is the motivation behind the LMCTS (Lookup Monte Carlo Tree Search) algorithm. As mentioned before, this cannot be implemented in the original algorithm since the optimal policy for the future time steps are unknown. LMCTS starts solving the problem from the end of the horizon, rather than the present time. First, we solve a one period problem and store the results to a lookup table. This lookup table will have the LMCTS solutions for all possible states at time n-1 for an n-period problem. Next we proceed to solve the two period problem. Now, we need to follow a rollout policy to complete our simulations. Instead of using a random rollout policy, or any other arbitrary rollout policy, we employ the results of the LMCTS lookup table as our rollout policy. We again store our results to our lookup table, which now includes the LMCTS solutions for both time n-1 and n-2. We continue implementing this algorithm until time 0.

One important caveat is that the LMCTS algorithm can only be implemented in a problem where the number of state-action pairs are tractable. Otherwise, the time it would take to calculate the lookup table and its size would explode as you move away from time n-1 towards time 0. In our multi-period portfolio problem, we luckily don't have such an issue for the no transaction cost case. When there are no transaction costs, the only thing that affects our portfolio is our belief of the regime. Hence, our lookup table will only consist of discretized beliefs and LMCTS solutions for the respective belief values. Note that, as we add new entries to the table, we only need to add a new set of LMCTS solutions for each discretized belief value. Therefore the size of the lookup table increases linearly as we move backward in time. 

If we were to solve the problem with transaction cost, we wouldn't be able to use LMCTS directly. In this case, the optimal portfolio at any time step depends on the previous time step's portfolio. This means that we would have to construct a table that stores the LMCTS value for every possible belief and portfolio duos. This would be practically impossible to do. In our two-step algorithm to solve the multi-period portfolio allocation problem with transaction costs, we implement the LMCTS solution for the no transaction cost case (and dyanmic programming for comparison), and let the neural network handle the transaction cost. Hence, the LMCTS algorithm is indeed viable in this algorithm pipeline.

Note that the LMCTS algortihm only modifies the simulation phase of the MCTS algorithm. It is compatible with any sort of selection, expansion and back-propagation methods.

\subsubsection{Kernel Regression and KR-UCT}

\paragraph{Kernel Regression} \leavevmode

Kernel Regression is a nonparametric technique that estimates the conditional expectation of a random variable. In 1964, Nadaraya and Watson (\citeauthor{nadaraya} \citeyear{nadaraya})(\citeauthor{watson} \citeyear{watson}) came up with the idea of estimating the expected value of a point based on a weighted average of the values of other points in the data set where the weights were inversely correlated with the distance between the points. The Nadaraya–Watson estimator for the expected value of a point is:

$$\mathbf{E}[y|x] = \frac{\sum_{i=0}^{n} K(x,x_i)*y_i}{\sum_{i=0}^{n}K(x,x_i)} \hspace{1cm}(11)$$

where $(x_i,y_i)$ represents the data and $K(.,.)$ is the kernel function. Different kernel functions can be used, but in this paper, a relatively popular function, the radial basis function (RBF), will be employed. The denominator of the Nadaraya–Watson estimator (11) is also called the kernel density which is a measure of relavant data to the point of interest:

$$W(x) = {\sum_{i=0}^{n}K(x,x_i)}  \hspace{1cm} (12)$$

The notations in this section  are chosen to be the same with (\citeauthor{Yee2016MonteCT} \citeyear{Yee2016MonteCT}).

\paragraph{KR-UCT in LMCTS and the Portfolio Allocation Problem} \leavevmode

\noindent The KR-UCT is an extension to the original UCT, which applies the kernel regression values $K(.,.)$ between the points for: information sharing between all actions under consideration, identification of actions outside of the initial candidates for further exploration, and selection of actions outside of the candidate set. It also uses a growing set of discrete actions, a version of the progressive widening idea, to handle continuous action spaces. We will now present the final version of the algorithm that we employ the solve our problem. This will be a version of the LMCTS algorithm where the selection and the expansion phases are governed by the KR-UCT function.

\textbf{Selection:} The idea of the selection phase is the same with the vanilla MCTS except the selection function. Instead of UCT, KR-UCT is employed. The definitions of $\mathbf{E}[v|a]$ and $W(a)$ can be seen below (\citeauthor{Yee2016MonteCT} \citeyear{Yee2016MonteCT}).

$$\mathbf{E}[v|a] = \frac{ \Sigma_{b \in A} K(a,b) \bar{v}_b n_b }{ \Sigma_{b \in A} K(a,b) n_b }  \hspace{1cm} (12)$$

$$W(a) = \Sigma_{b \in A} K(a,b) n_b  \hspace{1cm} (13)$$

This new KR-UCT function is basically UCT where $v$ is replaced with $\mathbf{E}[v|a]$ and $n$ is replaced with $W(a)$. The expected value of each action is now a function of the mean rewards of every single action node we have, weighted by their distance to the node in consideration. A closer node, where closeness is defined as having a relatively larger kernel function value, has a larger weight compared to a distant node and hence has a bigger effect on the expected value of the action node at hand. This allows the algorithm to share information between the nodes and gives us a better estimate of the potential value of the node.

\textbf{Expansion:}  As mentioned, this algorithm applies a variant of the progressive widening. Whether a new node will be added in the expansion phase or not is determined by a linear function of the number of visits to the node. In the case of an addition of a new node, we want a new node that is not too far away from our current best node for efficiency purposes, but we also want it to be relatively distant so that we can explore a new region in the action space. This trade-off is balanced by choosing the farthest node in terms of kernel density ($argmin W(a)$) out of the nodes which are at least $\tau$ close to the current best action ($K(action,a) > \tau$). $\tau$ is a hyper-parameter that can be tuned according to the data. There are some specific changes made to the algorithm in this part using our domain knowledge to get a faster algorithm. Domain knowledge in general consists of our knowledge of the problem at hand which allows us to come up with more efficient algorithms. We have identified two considerations.  First, it is expected for solutions of neighboring time steps to be similar to each other, for the same regime belief. To take advantage of this, the first node coming out of any node is manually coded to be the solution of the next timestep, which is retrieved from the lookup table of LMCTS. For the other expansions, we employ an approximation for the optimization described in this part as in (\citeauthor{Yee2016MonteCT} \citeyear{Yee2016MonteCT}). This is mainly done for computational efficiency. Instead of solving a fully fledged optimization problem, we determine a set of actions and select from these actions the one with minimal kernel density. The set of actions to be considered are the feasible portfolio allocations around the current best portfolio.

We have also implemented breadth wise expansion only, that is we are essentially employing trees with depth one at each time step. The point of building a deeper tree is to utilize better actions in the further time steps to more accurately measure the performance of the current actions in the tree. Since we are already using the optimal actions in the next time steps in LMCTS, there is no point in building a larger tree. The algorithm itself is totally compatible with a larger tree, but it would only cost unnecessary computational time.

{\centering
\includegraphics[width=1\columnwidth]{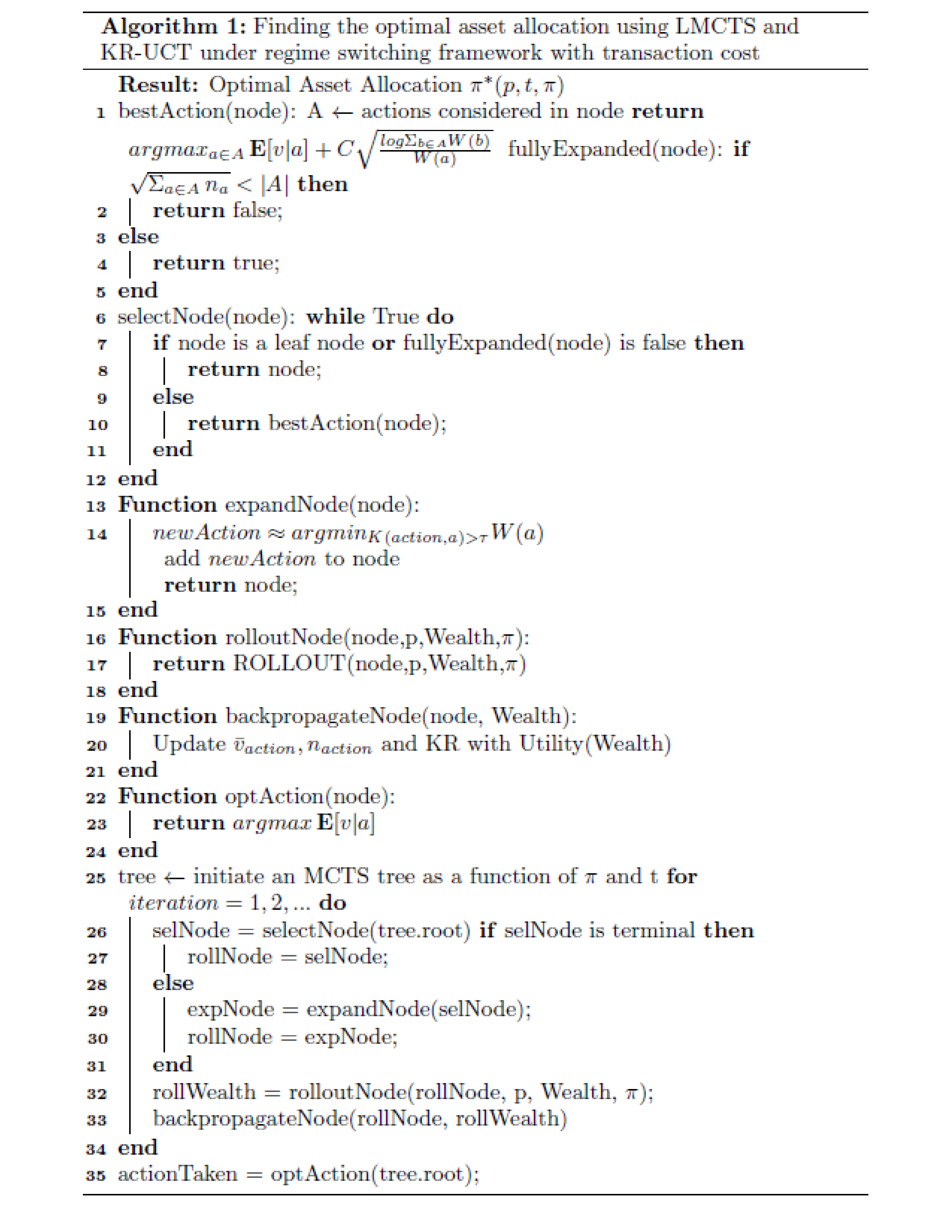}
}

\textbf{Simulation:} The main logic of the simulation phase is the same. We simulate until the end of our defined horizon and achieve a reward at the end. "ROLLOUT"  represents the rollout phase in the pseudo-code (line 16). The rollout policy followed here is actually the LMCTS rollout policy, which is the deterministic policy described in the previous sections. We are adopting a Markov switching framework for the regimes, so our belief of the current regime is constantly updated after we get new return values each period. This holds true for the original problem of the multi-asset portfolio allocation, but also for the individual simulations within the MCTS algorithm. The beliefs are updated:

$$p^{new}_k = \frac{pdf(r; \mu_k,\Sigma_k)*p_k} {\Sigma_{l=1}^N pdf(r; \mu_l,\Sigma_l)*p_l}  \hspace{1cm} (14)$$\;

and the wealth is updated for the original problem and for the simulations which is done as follows:

$$W^{new} = \Sigma_{i=0}^n \pi^i*(1+r^i) \hspace{1cm} (15)$$

\textbf{Backpropagation:} The backpropagation step is identical to the original MCTS, the relevant variables are updated along the traversed path using the reward obtained at the simulation step.

As mentioned, LMCTS uses information sharing across the nodes for each instance because we know that similar allocations should lead to similar returns. Another domain knowledge we have is that the optimal allocations in an asset at nearby time steps should not be too far away from each other. Depending on the size of the problem and the computational power at hand, a simulation based algorithm such as LMCTS may not fully converge, especially for time steps that are closer to the present time. We employ a commonly used signal filtering algorithm called Savitsky-Golay filter (\citeauthor{savitzky_golay_1964} \citeyear{savitzky_golay_1964}) for each asset independently across 50 timesteps to denoise the final result to give a more robust starting point for the neural network. The final results are normalized to ensure that we still obey the budget constraint.

\subsection{Neural Networks}
\subsubsection{General neural networks}
\begin{figure}[hbt!]
    \centering
    \includegraphics[width=0.6\textwidth]{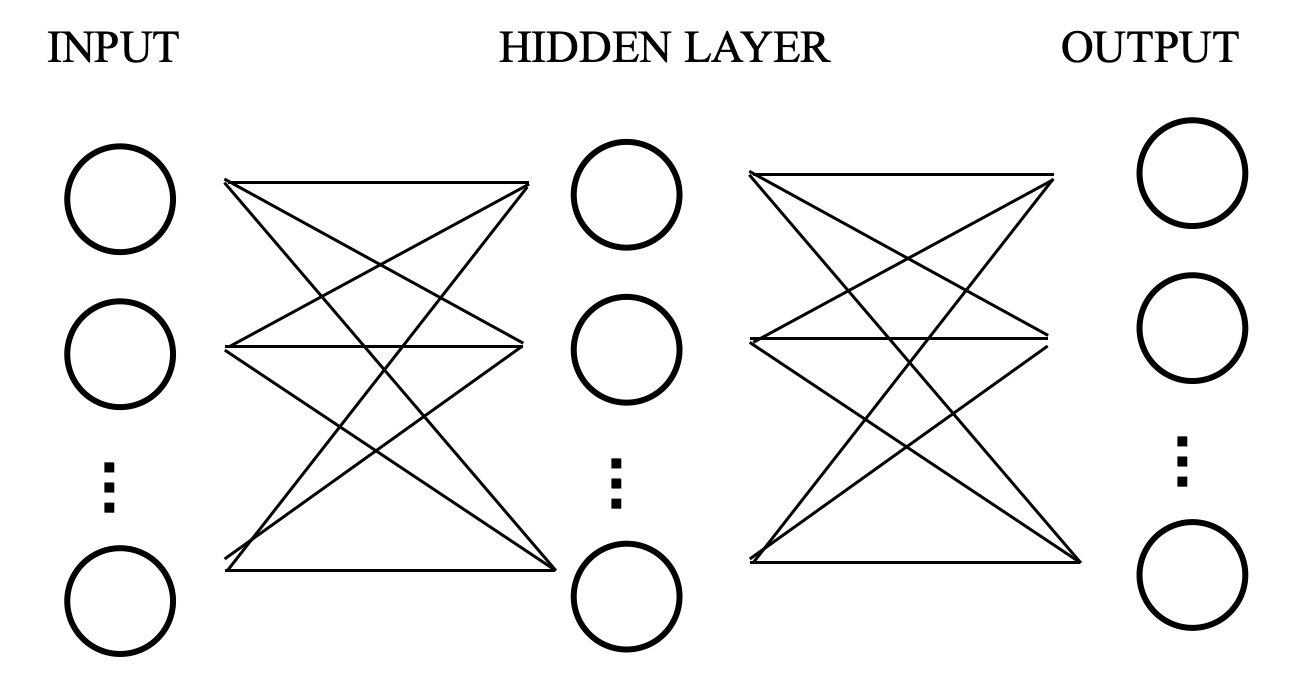}
    \caption{An artificial neural network with one hidden layer.}
    \label{fig:ann}
\end{figure}

Artificial neural networks, referred as neural networks in this paper, are inspired by the way biological neural systems process information. They are widely employed in predictive modeling and adaptive controlling. A neural network is based on a connected set of artificial neurons, where each connection is associated with a real number called weight that represents the relative strength of the connection. Figure \ref{fig:ann} exhibits an example of a simple neural network with one hidden layer. Such a graph that depicts the connection of the neurons is called the computational graph of a neural network. In this example, each neuron in the hidden layer is calculated as the weighted sum of the input layer (possibly with a bias term), passing through some nonlinear function called the activation function. Similarly, each neuron of the output layer is a weighted sum of the hidden layer, passing through an activation function.

When training, the neural network is fed with examples, and it adjusts the weights by gradient method to minimize the loss function. The process is called backpropagation. In each training step, the weights are updated according to the gradients of the loss function with respect to the weights as well as to the learning rate. Selecting learning rate is a non-trivial problem in training a neural network. On one hand, learning rate being too small causes the neural network to learn slowly and may result in unexpectedly long training time; on the other hand, if the learning rate is too large, the neural network may miss the global optimum.

\subsubsection{Recurrent neural networks}
A recurrent neural network (RNN) is a family of artificial neural networks, whose nodes form a directed connection. Unlike a feedforward neural network, the weights of a recurrent neural network is often shared across different time steps. The merit of sharing weights includes a) allowing for a temporal structure and b) successfully avoiding linearly growing number of parameters, which may lead to over-fitting or undesirably large number of required training samples. 

At each time step $t$ of a recurrent neural network, a hidden state $h_t$ is recorded. The hidden state is then fed into the next time step, possibly along with new information acquired at the next time step. Suppose there are $T$ time steps in a recurrent neural network, and the new information at each step is $x_0, x_1, ..., x_T$. The RNN evolves with
\begin{align*}
    h_t &= \theta(W_{hh}h_{t-1}, W_{xh}x_t),
\end{align*}

where $W_{hh}$ and $W_{xh}$ are the matrices of weights respectively, and $\theta$ is the activation function. Herein, we will employ recurrent neural network to learn the trading strategy of a multi-period asset allocation problem. In particular, when given complex problems, neural networks do not always converge to optimality in a fast and accurate way. However, it helps the neural networks to converge to global optima if the starting point is relatively close to the optimal solution. A good starting point does not only improves the performance overall, but  also shortens the time it takes to train the neural network. For our multi-period asset allocation problem involving transaction costs, we will take advantage of methods such as MCTS and dynamic program to find an approximate solution within reasonable running time, and then utilize neural networks to tune the allocation decision.

\subsubsection{Optimal no-trade zones via recurrent neural networks}

The optimal trading strategy under geometric Brownian motion(GBM) assets and proportional transaction cost is to form a no-trade zone (\citeauthor{davis} \citeyear{davis}). If we are in the no-trade zone, no action is required. If not, the agent is required to rebalance the portfolio to the closest point in the no trade zone. We are following the neural network structure in (\citeauthor{li2021} \citeyear{li2021}) to come up with no-trade zones. The no-trade zone is parameterized as follows:
$$u(p,\tau) = \pi(p, \tau) + f_{u}(p)$$
$$l(p,\tau) = \pi(p, \tau) - f_{l}(p)$$

where $p$ and $\tau$ represent our belief and time step respectively. $\pi(p, \tau)$ is the optimal solution under no transaction cost(embedded into the neural network as DP/LMCTS solution). $u(p,\tau)$ and $l(p,\tau)$ are the upper and lower no-trade zone boundaries. The goal of the neural network is to essentially learn  $f_{u}(p)$ and  $f_{l}(p)$.

\begin{figure}[hbt!]
    \centering
    \includegraphics[width=0.6\textwidth]{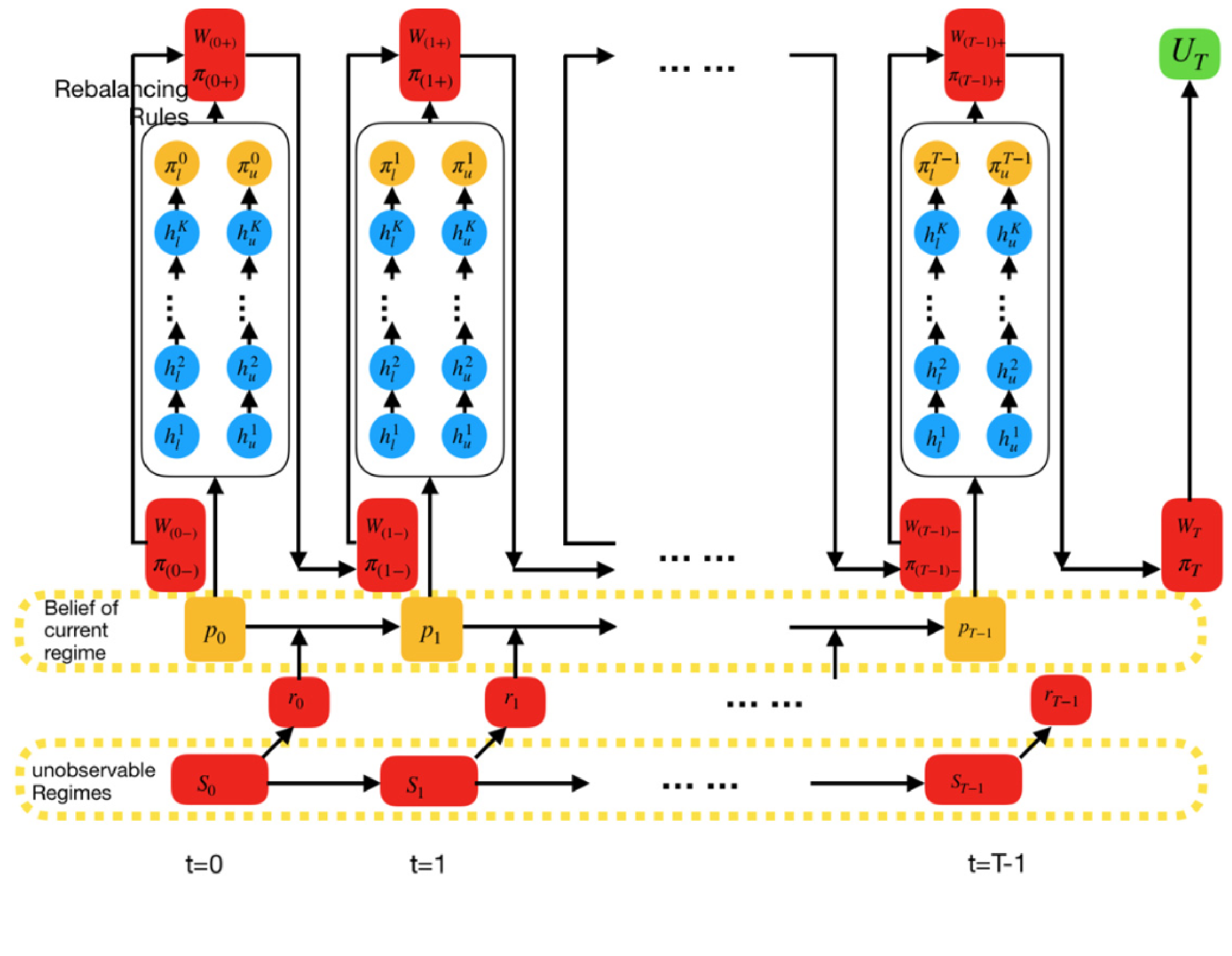}
    \caption{Computational graph of the neural network used.}
    \label{fig:comp_graph}
\end{figure}

The computational graph of the neural network can be seen in (Figure \ref{fig:comp_graph}) which is based on (\citeauthor{HE} \citeyear{HE}) and \citeauthor{mulvey2020} \citeyear{mulvey2020}) (2020). In the computational graph, $S_{t}$ is the unobservable real regime of market, $r_{t}$ is the vector of realized returns of the risky assets, $p_{t}$ is the agent's belief in the current market regime, $h_{l}$ and $h_{u}$ are the hidden layers in the neural network. $W_{(t^{-})}$ and $\pi_{(t^{-})}$ are the wealth and asset allocations at time t respectively before rebalancing. $\pi_{(t^{+})}$ is chosen based on the no-trade zone and the wealth after rebalancing is calculated as $W_{(t^{+})} = W_{(t^{-})} - cW_{(t^{+})}
\| \pi_{(t^{+})} - \pi_{(t^{-})} \|_{1}$. Finally, the loss function of the neural network is $-E[U(W_{T}]$. We encourage readers to refer to (\citeauthor{li2021} \citeyear{li2021}) to have a deeper understanding of the neural network structure and no-trade zones.

\subsubsection{Training neural network to maximize the probability of reaching goal}
We train the weights of neural networks via gradient propagation. When the objective is a CRRA utility function, there is a natural non-zero derivative of the objective with respect to weights. On the other hand, when the objective is the probability of reaching goal, the slope is zero almost everywhere. In addition, since transaction costs deteriorate the chance of reaching goal in a non-continuous manner, the choice of starting point is less obvious in this case. In this subsection, we provide a strategy to tackle the issues associated with this objective function. 

To maximize the probability of reaching a given goal, we choose to employ the DP/MCTS solution for CRRA utility under zero transaction costs as the starting point to feed into neural networks. The reason is two-fold: First, to maximize directly on probability, the current wealth needs to be considered in state space and therefore slows down the calculation of starting point. Secondly, the allocation strategy gets suboptimal as transaction costs are considered, and in particular, one shall almost always invest more heavily in early stages because the transaction costs consumes part of the profit and diminishes the probability of reaching goal. This makes the solution for maximizing probability deviate from the center of no-trade zone once transaction costs are added.

\begin{figure}[hbt!]
    \centering
    \includegraphics[width=0.6\textwidth]{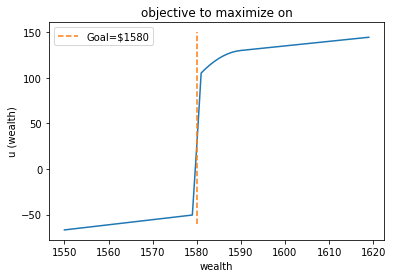}
    \caption{The approximated objective to use in neural network, when goal = \$1580.}
    \label{fig:obj_goal}
\end{figure}

To address the issue that current wealth would affect the allocation decision, we feed the current wealth along with the regime estimation into the neural networks. In addition, an extra neural network is placed to adjust the starting solution so that the no-trade zone is centered at the adjusted solution instead of the CRRA solution. Recall that the neural network is trained based on gradient propagation, whereas the objective of probability of reaching goal has gradient zero almost everywhere. To overcome this issue, we replace the objective with an approximated function who has positive slope everywhere (Figure \ref{fig:obj_goal}). Please refer to Section 4 to the empricial problem setup to have a better understanding of (Figure \ref{fig:obj_goal}).

\subsection{Dynamic Program}
Dynamic programming is a algorithmic technique for solving optimization problems, where the original problem is break down to simpler subproblems, and the solutions to subproblems are utilized for solving the original problem. It is widely used in sequential problems, where subproblems are often well-defined by its nature. To fully describe a dynamic program, one defines the state space that includes all information needed for decision making, the action space that contains all possible actions to take, a transition function that tells the outcome of certain actions under some state, and a value function that quantifies the goodness of an action under a state. For example, in our setting, a value function returns the expected terminal utility of taking an action given the current market environment.

It is a natural algorithm for our multi-period asset allocation problem, if one has sufficient computational power to solve for the whole system. However, when the size of state space or that of the action space grows, dynamic program faces the curse of dimensionality, meaning that the running time would grow exponentially with the complexity of the problem. In our allocation example, if there are no transaction costs, we may simplify the state space to a two dimensional space that includes the probabilistic estimation of underlying regime and the time until horizon. The prices of risky assets does not need to be included in the state space when CRRA utility is considered, as one can easily scale the wealth. On the other hand, if transaction costs are taken into consideration in a dynamic program, an allocation decision must depend on the current weights in each asset, and therefore the state space grows with the number of risky assets. Practically we find it intractable to directly apply a vanilla dynamic program to the multi-period asset allocation problem involving transaction costs. Instead, we propose it to be one of the methods that finds allocation decisions under zero transaction costs, and could help us to gain an advance starting point for neural networks.

\section{Empirical Results}

The problem setting for the empirical experiments is defined in the previous sections. We have 11 risky assets and 1 risk-free asset, where the mean and the covariance matrix of the assets were determined by the historical values of real assets. The parameters of the hidden Markov model are calibrated on the following stocks based on weekly returns from January 1, 2000 to December 14, 2019: AAPL, IBM, AMZN, JPM, BRK, GE, JNJ, NSRGY, T, XOM, and WFC. To enable simulations and illustrate computational advantages on average, we assume the market dynamic is stationary and that the parameters are kept fixed over the 50-week horizon. We start at an arbitrary wealth of \$1000 and the objective function for maximizing terminal utility is CRRA utility with risk aversion parameter $\gamma=-1$.

The neural network and the dynamic programming follows the same specifications in (\citeauthor{li2021} \citeyear{li2021}). For the LMCTS, the exploration/exploitation tradeoff parameter is taken to be 5. The belief probabilities are discretized to be multiples of 5\%.  For the expansion phase of the LMCTS, the candidate portfolios that we consider are the feasible portfolios that deviate 5\%, 10\%, and 20\% away in a single asset from the current best portfolio. We have used a first order Savitsky Golay filter with a window size of 11 to smoothen the results. 10000 simulations are used for each belief/time pair and the reward at each iteration is calculated as the mean of 50000 return paths. A set of return paths starting from each possible belief state is simulated beforehand to expedite the computations.

\subsection{Terminal utility maximization without shorting}
First, we examine at the case where shorting the assets is not allowed. Note that being able to short an asset causes the state space for the asset allocation to increase immensely for the LMCTS algorithm. The reason for the increase in complexity is that since we discretize the asset allocation space for the LMCTS, the budget constraint for the no shorting scenario helps us to shrink the state space considerably, which does not quite work effectively for the scenario where shorting is allowed.

A lower bound of the optimal expected utility is provided with the optimal trading strategy under zero transaction costs. We call the strategy ``adjusted DP'' because it deals with the no-shorting constraint with a penalty term of shorting in the objective function. This strategy is learned by a dynamic program, and involves rebalancing at each time period which leads to high transaction costs. 

\begin{figure}[h]
    \centering
    \includegraphics[width=0.6\textwidth]{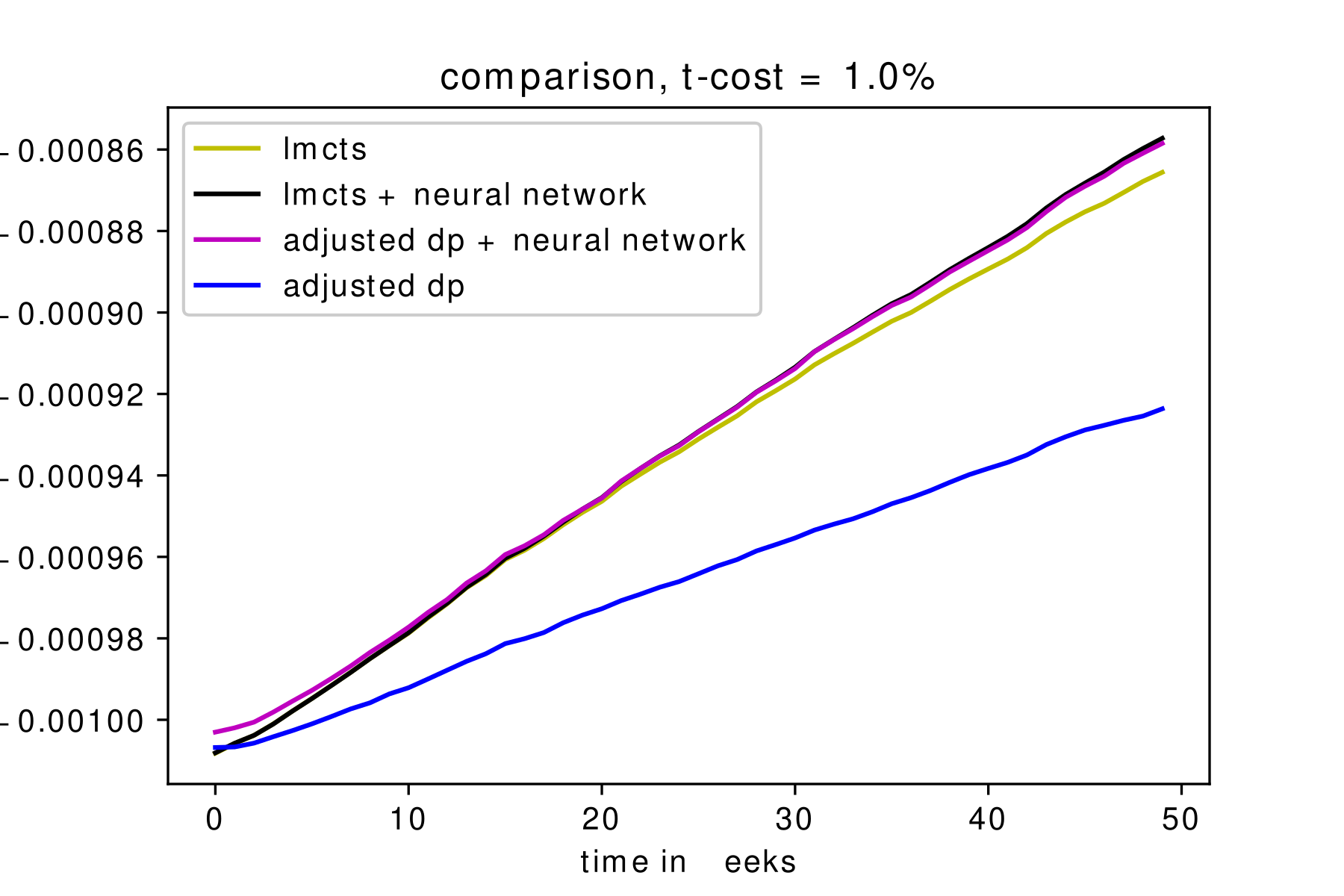}
    \caption{Utility comparison of different methods without shorting.}
    \label{fig:comparison_nolev}
\end{figure}

The summary of the results can be seen in (Figure \ref{fig:comparison_nolev}). The utility path of numerous simulations have been gathered to produce the comparison graph. We see that LMCTS+NN and DP+NN are the best performing algorithms overall, which was expected. Remember that LMCTS and DP are actually solving the question for zero transaction cost and their results are only used for starting points for the neural network. We see that the LMCTS result significantly outperforms the adjusted dp result without the neural networks. This translates into the LMCTS+NN very slightly outperforming DP+NN, which definitely shows us that LMCTS is the better algorithm to use in the case of a constrained state space.

\subsection{Terminal utility maximization with shorting}

\begin{figure}[h]
    \centering
    \includegraphics[width=0.6\textwidth]{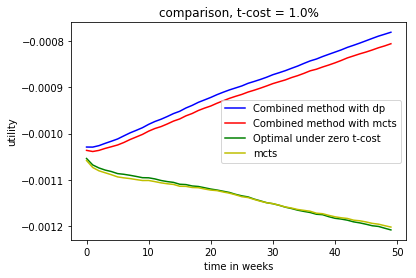}
    \caption{Utility comparison of different methods with shorting.}
    \label{fig:utility_comparison_with_shorting}
\end{figure}

Now we explore the case where shorting the assets are allowed. For each asset, we are allowing an allocation between -100\% and 100\% of our total capital. The results can be seen in (Figure \ref{fig:utility_comparison_with_shorting}).

In the case of shorting, we again see that LMCTS has comparable performance to the optimal solution under zero transaction cost(DP). Of course, one of the reasons for this is that the LMCTS algorithm works with a discretized space, so it rebalances less than the DP, which results in less incurred transaction cost. DP+NN does perform better than LMCTS+NN though after the neural network deals with the transaction costs.\\
The main difference between shorting and no-shorting experiments is that in the latter case, the state space is significantly larger. Not only we can short each asset, but once we do so, we end up with a higher capital that can be allocated into the other assets. Similarly, longing asset creates more combinations of shorting for other asset. It is still promising to see that LMCTS gives a very competitive good result under a setting which is challenging for its structure.

\subsection{Maximizing the probability of reaching a goal}
Another subject of interest is the ability of the algorithms to discover good solutions under different objective functions. In the previous sections, maximizing the terminal utility was our objective function. Now we examine algorithmic performance when the objective function is maximizing the probability of reaching a goal, a non-convex optimization model.

\begin{figure}[hbt!]
    \centering
    \includegraphics[width=0.6\textwidth]{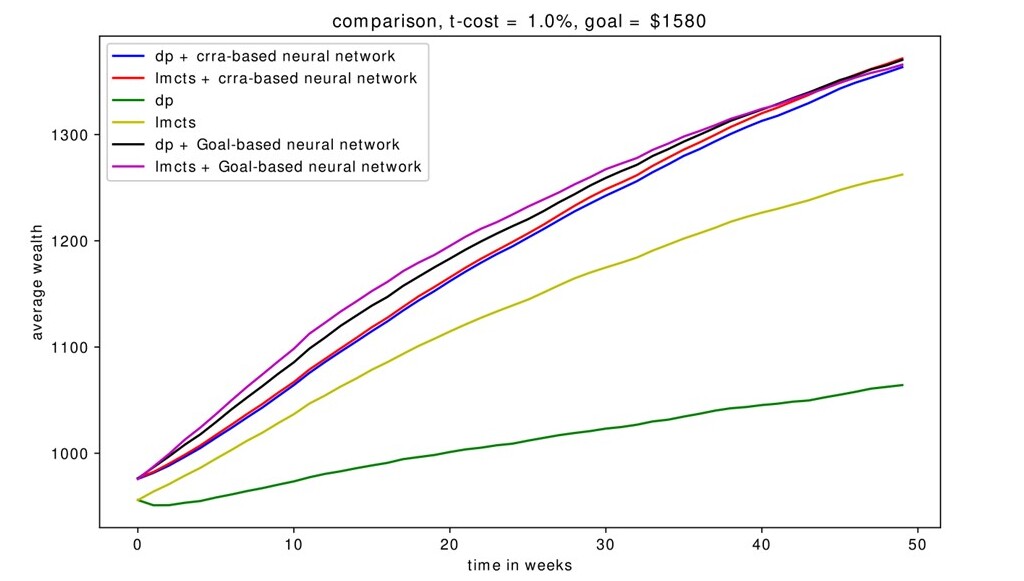}
    \caption{Comparison of different methods for probability of reaching a goal as the objective function. The goal is set as \$1580.}
    \label{fig:goal_prob_reaching_goal}
\end{figure}

\begin{figure}[hbt!]
    \centering
    \includegraphics[width=0.7\textwidth]{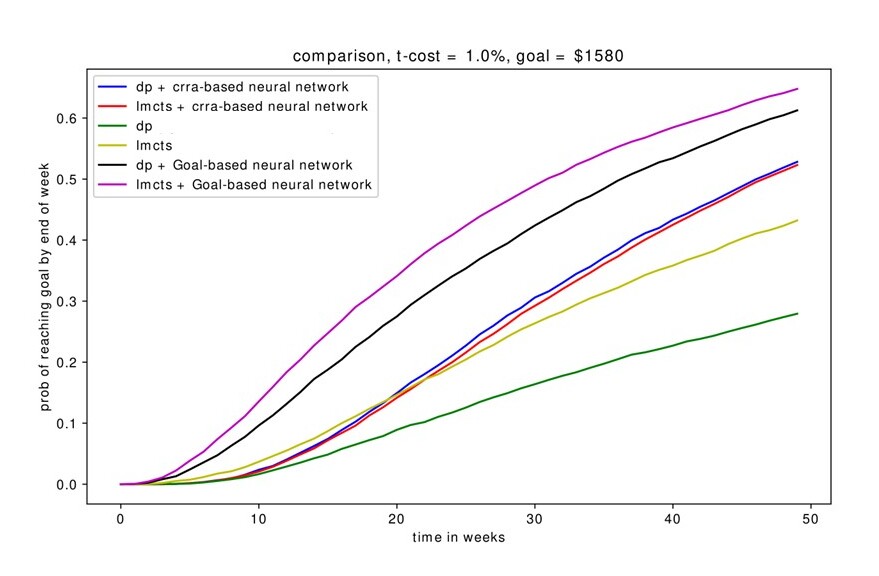}
    \caption{Wealth comparison of different methods for probability of reaching a goal as the objective function. The goal is set as \$1580.}
    \label{fig:goal_avg_wealth}
\end{figure}

This time we present two different graphs when the goal is \$1580 without shorting, one for the probability of reaching a goal (Figure \ref{fig:goal_prob_reaching_goal}), which is the main objective,  and one for the wealth paths of the algorithms (Figure \ref{fig:goal_avg_wealth}). Similar to previous sections, DP+NN and LMCTS+NN perform significantly better than the standalone algorithms, with goal-based neural networks (which optimizes on probability of reaching goal) outperforming CRRA-based neural networks (which optimizes on the CRRA utility). We see that the LMCTS+NN provides the best performance overall.

\subsection{The effect of transaction cost}

\begin{figure}[h]
    \centering
    \includegraphics[width=0.6\textwidth]{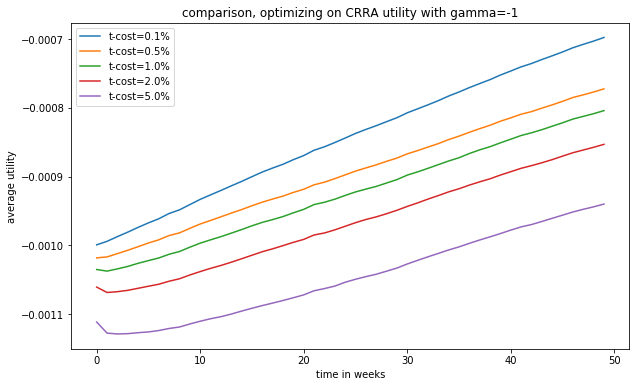}
    \caption{Wealth comparison of MCTS+NN for different transaction costs for probability of reaching a goal as the objective function}
    \label{fig:comparison_tcost_utility_mcts}
\end{figure}

\begin{figure}[h]
    \centering
    \includegraphics[width=0.6\textwidth]{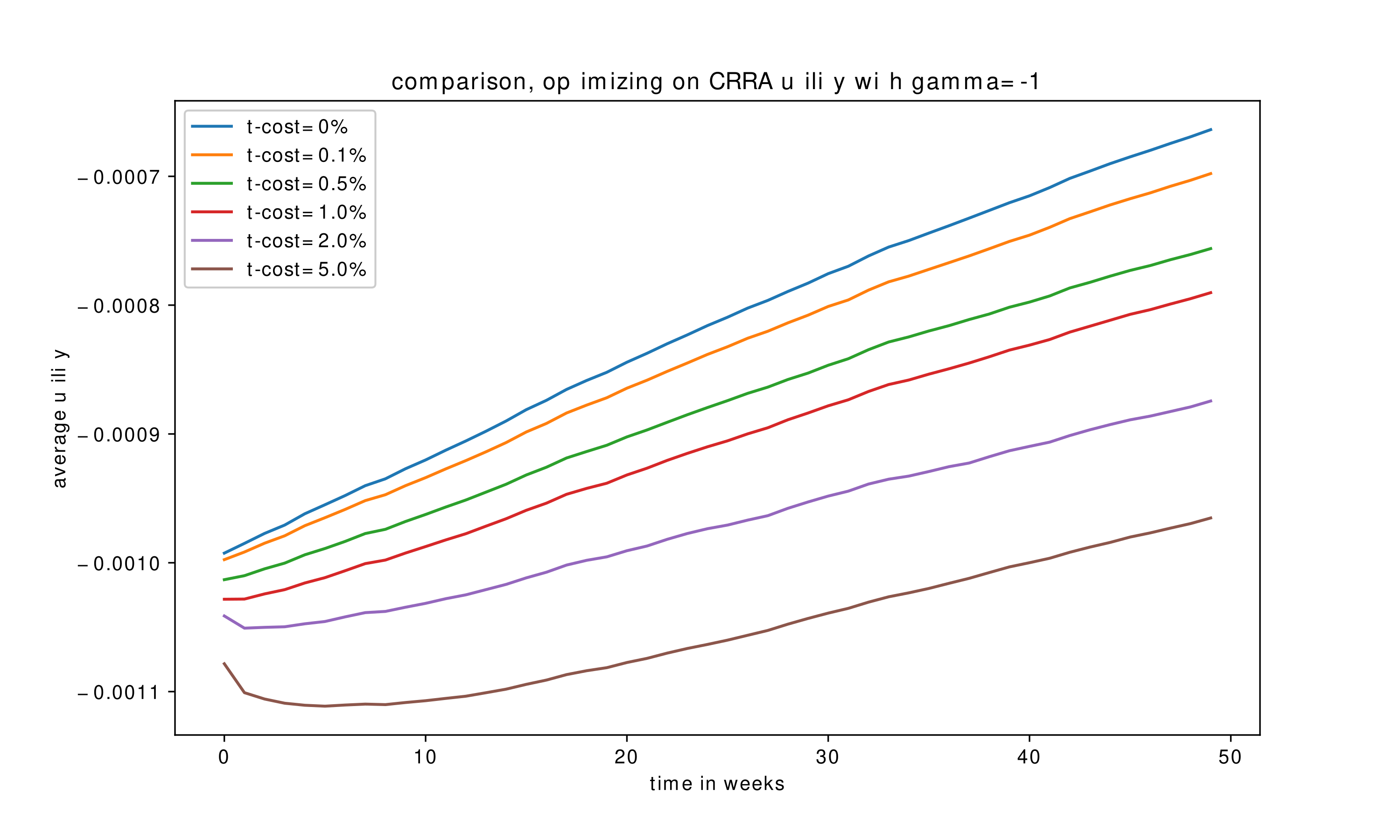}
    \caption{Wealth comparison of DP+NN for different transaction costs for probability of reaching a goal as the objective function}
    \label{fig:comparison_tcost_utility}
\end{figure}

Another significant parameter is transaction costs. The performance of the algorithms under different linear transaction costs can be seen in (Figure \ref{fig:comparison_tcost_utility_mcts}) and (Figure \ref{fig:comparison_tcost_utility}), for MCTS+NN and DP+NN respectively.

We see that as the transaction cost decreases, we can reach higher terminal utility values at the end of the horizon,as expected. For higher transaction cost values, there is a significant decrease in the utility at the beginning of the time horizon. The algorithms render an early adjustment at the expense of some wealth at the beginning of the horizon to reach a favorable portfolio. Both algorithms recover from the initial loss rapidly and reach a higher terminal value at the end.

\section{Conclusions}

The MCTS is an algorithm that can be employed in real-world finance problems. The computational advantage of combining neural networks with a MCTS or DP algorithm becomes evident as the size of the problem grows. When traditional algorithms suffer from the curse of dimensionality, the combined methods strive to offer an efficient way in finding solutions. In this paper, we explored a multi-period financial portfolio optimization problem under a regime-switching hidden Markov model and shown that the MCTS, more specifically LMCTS, algorithm yields sufficient starting points for a deep neural network algorithm to achieve excellent results. In select cases, the results are  better than the DP+NN duo which serves as a viable benchmark to test performance. A noticeable advantage of the algorithm is its adaptability due to the simulation environment, which makes it a general purpose tool. A challenge of the MCTS algorithm is exponential growth in run time with regard to the structure of the model. This suggests that alternative models can be exploited going forward. The introduced LMCTS algorithm together with the KR-UCT function aims to reduce this dependency. By using a lookup table and a more efficient selection/expansion function LMCTS manages competitive results in larger state spaces, as compared to its dynamic programming benchmark.

\newpage 

\section{Future Work}

As mentioned, we have used the CRRA utility function solutions of the LMCTS and DP algorithms as the starting points for the case of maximizing the probability of reaching a goal objective function. LMCTS is in fact able to handle such an objective function by itself, a reward of 1 if we reach the goal and 0 if not, but we haven't used it for two reasons. Firstly, the lookup table would have to include an extra dimension consisting of discretized possible wealth values which would take a lot of computational time, and secondly our neural network isn't compatible with such a starting point by design. It would be of interest to experiment with diferent kinds of non-convex functions such as probability of reaching a goal with LMCTS. Another extension could be to constrain the state space, for example having no allocations over 20\% in any asset. The LMCTS would handle this by removing the infeasible solutions from the candidate actions in the expansion phase.

\newpage
\bibliographystyle{plainnat}
\bibliography{reference}

\end{document}